\documentclass{iopart}
\usepackage{iopams,amsfonts,amssymb,amsthm} 
\usepackage{geometry}                
\usepackage{graphicx}
\usepackage{rawfonts}
\usepackage{amssymb}
\usepackage{epstopdf}
\usepackage[usenames,dvipsnames,svgnames]{xcolor}
\input prepictex.tex
\input pictex.tex
\input postpictex.tex

\newtheorem{theo}{Theorem}
\newtheorem{lemm}{Lemma}

\newtheorem{remark}{Remark}

\def\Pr{\noindent \emph{Proof: }}
\def\qed{$\Box$}

\def\nor{\normalsize}

\def\IntN{\mathbb{Z}}
\def\sfrac#1#2{\hbox{\nor $\frac{#1}{#2}$}}

\def\shalf{{\sfrac{1}{2}}}

\def\squarter{{\sfrac{1}{4}}}

\begin{document}

\title{Self-avoiding walks adsorbed at a surface and pulled at their mid-point}
\author{E J Janse van Rensburg$^*$ and S G  Whittington$^\dagger$ }
\address{
{}$^*$Department of Mathematics \& Statistics, York University, M3J 1P3, Toronto, Canada \\
{}$^\dagger$Department of Chemistry, University of Toronto, M5S 3H6, Toronto, Canada 
}

\begin{abstract}
We consider a self-avoiding walk on the $d$-dimensional hypercubic lattice, terminally attached
to an impenetrable hyperplane at which it can adsorb.  When a force is applied the walk can be pulled off
the surface and we consider the situation where the force is applied at the middle vertex of the walk.  We
show that the temperature dependence of the critical force required for desorption differs from the
corresponding value when the force is applied at the end-point of the walk.   This is of interest in single 
molecule pulling experiments since it shows that the required force can depend on where the force is
applied.  We also briefly consider the situation when the force is applied at other interior vertices
of the walk.

\end{abstract}

\pacs{82.35.Lr,82.35.Gh,61.25.Hq}
\ams{82B41, 82B80, 65C05}
\submitto{J Phys A}
\maketitle

\section{Introduction}
\label{sec:Introduction}

Self-avoiding walks are the standard model of the configurational properties of long
linear polymers in dilute solution \cite{Rensburg2015,MadrasSlade}.   The situation can be adapted
to model the adsorption of linear polymers at an impenetrable 
surface \cite{HTW,Rensburg1998,RensburgRechnitzer,RychlewskiJSP} 
and the general features of the adsorption behaviour are now quite well
understood.  
With the invention of micro-manipulation techniques such as atomic force 
microscopy (AFM) and optical tweezers that allow individual polymer molecules to be 
pulled \cite{Haupt1999,Zhang2003} there has been renewed interest in how polymers 
respond to a force and, specifically, how self-avoiding walk models of polymers
respond to a force
\cite{Beaton2015,GuttmannLawler,IoffeVelenik,IoffeVelenik2010,Rensburg2009,Rensburg2016a}.
There has also been some work on how lattice polygons (a model of ring polymers)
respond to a force \cite{GuttmannLawler,Rensburg2008a,Rensburg2008b}.

In this paper we shall be concerned with self-avoiding walks adsorbed at a surface and pulled 
off the surface (\emph{i.e.} desorbed) by the application of a force.  The case that has received most attention
is a self-avoiding walk on the $d$-dimensional hypercubic lattice $\IntN^d$,
attached at one end point to an impenetrable surface at which it can adsorb.  A force,
normal to the surface, is applied at the other vertex of degree 1 (\emph{i.e.} at the other end point
of the walk) and this force is increased until the walk desorbs from the surface 
\cite{Guttmann2014,Rensburg2013,Krawczyk2005,Krawczyk2004,Mishra2005}.  At a particular temperature $T$
(below the critical temperature for adsorption) there is a critical value of the force, $f_c(T)$.  If
the applied force is less than $f_c(T)$ the walk is adsorbed while if the force is greater than
$f_c(T)$ the walk is desorbed into a ballistic phase.  If $d \ge 3$ then the 
force-temperature curve is reentrant, \emph{i.e.} the critical force initially increases as the 
temperature is increased at low temperature \cite{Rensburg2013,Krawczyk2005,Krawczyk2004,Mishra2005}.
The walk has entropy in the adsorbed state and this entropy is lost at low temperature 
when the walk is pulled off the surface.  The reentrance is associated with the force 
required to compensate for this entropy loss.
See also \cite{Skvortsov2009} and \cite{Binder2012} for related work.
In two dimensions the critical force is a monotone decreasing function of the temperature
\cite{Guttmann2014,Krawczyk2005,Krawczyk2004,Mishra2005}.  For all $d \ge 2$ the phase 
transition from the adsorbed to the ballistic phase is first order \cite{Guttmann2014}.

In an AFM experiment, unless special precautions are taken, the AFM tip can be 
in contact with different monomers, not just the last monomer.  Consequently it is
natural to ask how the behaviour depends on where on the polymer the force is 
being applied.  Apart from the case discussed above where the force is applied
at the last monomer the only situation that has been 
studied \cite{GuttmannLawler,Rensburg2016a,Rensburg2016b} is as follows.  Suppose
that we imagine a plane, parallel to the adsorbing plane, containing the monomers that 
are furthest away from the adsorbing plane, and apply the force either to pull this plane
away from or push it towards the adsorbing surface.  We can think of the force as being
conjugate to the span of the polymer in the direction normal to the adsorbing plane.
Beaton \emph{et al} \cite{GuttmannLawler} looked at the situation where there is no 
interaction with the adsorbing plane (except that it is impenetrable) and considered
pushing towards this plane.  They used ideas from SLE to make some 
predictions in two dimensions, and checked these by exact enumeration and 
series analysis.  They discovered
interesting sub-exponential behaviour that causes slow convergence to the limiting
behaviour.  The limiting behaviour when there is a surface interaction
and a force has also been investigated \cite{Rensburg2016b}.

\begin{figure}[h] 
\begin{minipage}[t]{0.475\linewidth}
\flushleft
\input figure1.tex
\vspace{5mm}
\caption{An adsorbing polymer pulled at its midpoint by a force $f$ in the
vertical direction.  Monomers in the polymer adsorb in the hard wall with activity $a$. }
\label{figure1}
    \end{minipage}%
\begin{minipage}[t]{0.475\linewidth}
        \flushright
\input figure2.tex
\vspace{5mm}
\caption{ An adsorbing walk pulled at its midpoint by a force $f$ in the
vertical direction.  Monomers in the walk interact with the hard wall with activity $a$. }
\label{figure2}
    \end{minipage}%
\end{figure}

In this paper we are looking at the situation illustrated in figure \ref{figure1},
where the adsorbing polymer is pulled in its midpoint from the adsorbing surface.
This is modelled by a self-avoiding walk as shown in figure \ref{figure2}:  The 
force is applied at the midpoint of the walk, and vertices of the walk interact
with the adsorbing surface with activity $a$.

\section{Some notation and a brief review}
\label{sec:notation}

Consider the $d$-dimensional hypercubic lattice $\IntN^d$ and attach the obvious coordinate 
system $(x_1,x_2,\ldots x_d)$ so that each vertex has integer coordinates.  The hyperplane
$x_d = 0$ will be the distinguished plane at which adsorption can occur.  A \emph{positive
walk} is a self-avoiding walk that starts at the origin and has $x_d \ge 0$ for all vertices
of the walk, so that it is confined to be in or on one side of $x_d=0$.  Let $c_n^+(v,h)$ be the 
number of $n$-edge positive walks with $v+1$ vertices in $x_d=0$ and with the $x_d$-coordinate
of the last vertex equal to $h$.  We call $h$ the \emph{height} of the last vertex and we say that the 
walk has $v$ \emph{visits}.  Define the partition function
\begin{equation}
C_n^+(a,y) = \sum_{v,h} c_n^+(v,h)\, a^vy^h.
\end{equation}
We can write $a = e^{-\epsilon/k_B T}$ and $y=e^{f/k_BT}$ where $\epsilon$ is the 
energy associated with a vertex in the surface, $k_B$ is Boltzmann's constant, $T$ is
the absolute temperature and $f$ is the force applied to the last vertex, measured in energy units.
For adsorption to occur $\epsilon$ must be negative so $a > 1$.  A force directed away from the surface 
corresponds to $ f > 0$ or $y>1$.
It is known \cite{Rensburg2013} that the limit
$\lim_{n\to\infty} \sfrac{1}{n} \log C_n^+(a,y) \equiv \psi (a,y)$
exists for all $a$ and $y$.  We shall write $\psi(a,1) = \kappa(a)$ and $\psi(1,y) = \lambda(y)$.
$\kappa(a)$ is the free energy of an adsorbing walk in the absence of a force \cite{HTW}
and $\lambda(y)$ is the free energy of a walk subject to a force but not interacting with the surface
\cite{Beaton2015}.  $\kappa(a)$ is a convex function of $\log a$ and there is a critical value
of $a$, $a_c > 1$, such that $\kappa(a) = \log \mu_d$ when $a \le a_c$ and $\kappa(a) >
\log \mu_d$ when $a > a_c$ \cite{HTW}.  Here $\mu_d$ is the growth constant of self-avoiding walks
on $\IntN^d$ \cite{Hammersley1957}.  Similarly $\lambda(y)$ is a convex function of $\log y$ \cite{Rensburg2009},
equal to $\log \mu_d$ when $y \le 1$ and greater than $\log \mu_d$ when $ y > 1$ \cite{Beaton2015}.
See also \cite{IoffeVelenik,IoffeVelenik2010}.   We know that \cite{Rensburg2013}
\begin{equation}
\psi(a,y) = \max [\kappa(a),\lambda(y)]
\end{equation}
so, when $a>a_c$ and $y>1$,  there is a phase boundary in the $(a,y)$-plane determined by the solution
of the equation $\kappa(a) = \lambda(y)$, between
an adsorbed phase and a ballistic phase.  This phase transition is first order \cite{Guttmann2014}.

If the walk is pulled or pushed at its top plane then we need to keep track of the span of the 
walk in the $x_d$-direction.  Let $c_n(v,s)$ be the number of $n$-edge positive walks with
$v+1$ vertices in $x_d=0$ and with span in the $x_d$-direction equal to $s$.  Define
the partition function 
\begin{equation}
C_n(a,y) = \sum_{v,s} c_n(v,s)\, a^v y^s.
\end{equation}
The limit $\lim_{n\to\infty} \sfrac{1}{n} \log C_n(a,y)$ exists and is equal to $\psi(a,y)$
\cite{Rensburg2016b} so, in the infinite $n$ limit, the free energy is identical to the 
free energy when the force is applied at the last vertex.  There are, however, major differences
in the finite size behaviour \cite{GuttmannLawler,Rensburg2009}.

Are there situations where the location where the force is applied leads to different behaviour?
In this paper we shall show that there are.  We focus on the effect of applying the force at the 
middle vertex of the walk, although we shall show in Section \ref{sec:SAWinterior}
that, in some circumstances, our results generalize in a natural way to pulling at other interior vertices,
while in other circumstances there is an additional phase in the phase diagram. 

Number the vertices of the walk $0, 1, 2, \ldots n$.
We define the \emph{middle vertex} to be the vertex numbered $\shalf n$ if $n$ is even and 
$\shalf(n-1)$ if $n$ is odd.  Let $w_n(v,h)$ be the number of $n$-edge positive
walks with $v+1$ vertices in $x_d = 0$ and with the $x_d$-coordinate of the middle
vertex equal to $h$.  We call $h$ the \emph{height} of the middle vertex.   Define the 
partition function 
\begin{equation}
W_n(a,y) = \sum_{v,h} w_n(v,h)\, a^vy^h.
\label{eqn:PFmiddle}
\end{equation}
We shall show that the limit 
\begin{equation}
\lim_{n\to\infty} \sfrac{1}{n} \log W_n(a,y) \equiv \phi(a,y)
\end{equation}
exists for all $a$ and $y$, and explore its relation to $\psi(a,y)$.  In particular we shall
show that the two free energies are not equal in some regions of the $(a,y)$-plane.
In fact, as we shall see, the two free energies are equal in the \emph{free phase} when
$0 \le a \le a_c$ and $0 \le y \le 1$ (see Section \ref{sec:SAWpushed}), and in the adsorbed phase, 
but not in the ballistic phase (see Section \ref{sec:SAWpull}).  Consequently the phase 
boundary between the adsorbed and ballistic phases
is different when the walk is pulled at the middle and at the end vertex.

A \emph{bridge} is a positive walk with the extra conditions that
\begin{enumerate}
\item 
The first edge is in the $x_d$-direction, and
\item
The $x_d$-coordinate of the last vertex is at least as large as that of any
other vertex.
\end{enumerate}
Let $b_n(h)$ be the number of $n$-edge bridges with the $x_d$-coordinate of the last vertex being 
$h$, and define the partition function as $B_n(y) = \sum_h b_n(h) y^h$.  
Then $\lim_{n\to\infty} \sfrac{1}{n} \log B_n(y) = \lambda (y)$ \cite{Rensburg2016a}.

Define a \emph{loop} to be a positive
walk with both vertices of degree 1 in $x_d=0$.  Let
$l_n(v,s)$ be the number of $n$-edge loops with 
$v+1$ vertices in $x_d=0$ and with span in the $x_d$-direction equal to $s$.
Write $L_n(a,y)=\sum_{v,s} l_n(v,s)\, a^v y^s$ for the partition function of 
loops with $y$ conjugate to the span in the $x_d$-direction. 
Then \cite{HTW} $\lim_{n\to\infty} \sfrac{1}{n} \log L_n(a,1) = \kappa(a)$.
Since the end vertices of a walk can be somewhat inaccessible we shall often find it useful
to work with unfolded walks \cite{HammersleyWelsh} and we recall some results about 
unfolded objects of various types.  
Write $x_i(j)$ for the $i$th coordinate of the $j$th vertex of an $n$-edge walk or loop, $1 \le i \le d$,
$0 \le j \le n$.
A loop is \emph{unfolded} if $x_1(0) \le x_1(j) < x_1(n)$ for all $0 < j < n$ and we write 
$L_n^{\ddagger}(a,y)$ for the partition function of unfolded loops (with $y$ conjugate to the span
in the $x_d$-direction).  In a similar way we write $W_n^{\ddagger}(a,y)$ for the partition function
of unfolded walks pulled at their mid-point (with $y$ conjugate to the height of the 
middle vertex) and $C_n^{\ddagger}(a,y)$ 
for the partition function of unfolded positive walks (with $y$ conjugate to the height of the last vertex).
For these three cases we have \cite{HTW,HammersleyWelsh},
\begin{eqnarray}
L_n^{\ddagger}(a,y) & \le & L_n(a,y) \le e^{O(\sqrt n)} L_n^{\ddagger}(a,y); \nonumber \\
C_n^{\ddagger}(a,y) & \le & C_n^+(a,y) \le e^{O(\sqrt n)} C_n^{\ddagger}(a,y);  \\
W_n^{\ddagger}(a,y) & \le & W_n(a,y) \le e^{O(\sqrt n)} W_n^{\ddagger}(a,y). \nonumber
\end{eqnarray}
Hence $\lim_{n\to\infty} \sfrac{1}{n} \log L_n^{\ddagger}(a,y)
= \lim_{n\to\infty} \sfrac{1}{n} \log L_n (a,y)$, and similarly for $C_n(a,y)$ and
$W_n(a,y)$.  In a similar way we write $B_n^{\ddagger}(y)$ for the partition function of unfolded bridges,
and 
\begin{equation}
B_n^{\ddagger}(y) \le B_n(y) \le e^{O(\sqrt n)} B_n^{\ddagger}(y).
\end{equation}

\section{Walks pushed towards the surface at their middle vertex}
\label{sec:SAWpushed}

In this section we consider the situation where the middle vertex is being 
pushed towards the surface.  That is, $f < 0$ or $y < 1$.  When $y=1$ 
\cite{HTW} we know that 
\begin{equation}
\lim_{n\to\infty} \sfrac{1}{n} \log L_n^{\ddagger}(a,1) 
= \lim_{n\to\infty} \sfrac{1}{n} \log C_n^{\ddagger}(a,1) = \kappa (a).
\end{equation}

\begin{figure}[t]
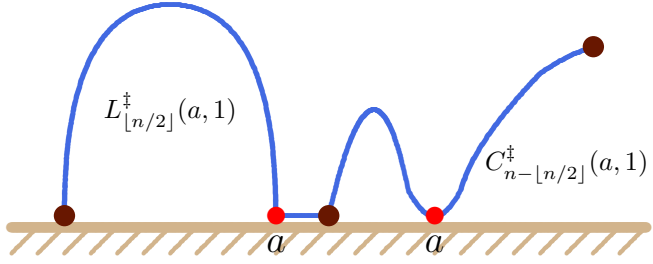

\input figure3.tex
\caption{Concatenating an unfolded adsorbing loop with an unfolded walk gives 
a lower bound on $W_n(a,0)$ (that is, the midpoint is at height zero).}
\label{figure3}
\end{figure}

\begin{theo}
For all $a > 0$ and $y \le 1$ the free energy of walks with the force applied at the middle 
vertex is equal to the free energy of walks with the force applied at the end vertex.  Moreover,
this free energy is independent of $y$.
That is, $\phi(a,y) = \psi(a,y) = \psi(a,1) = \kappa(a)$ for all $a>0$ when $y \le 1$.
\end{theo}
\Pr
When there is no force it is clear that $\phi (a,1) = \kappa (a) = \psi(a,1)$.  
Fix $y < 1$.  By monotonicity $W_n(a,0) \le W_n(a,y) \le W_n(a,1)$ and therefore
\begin{equation}
\limsup_{n\to \infty} \sfrac{1}{n} \log W_n(a,y) \le \kappa (a) = \psi(a,1).
\label{eqn:limsupW}
\end{equation}
To  get a bound in the other direction note that, for all $a>0$,
\begin{equation}
W_n(a,y) \ge W_n(a,0) \ge L_{\lfloor n/2 \rfloor}^{\ddagger}(a,1) \,
C_{n-\lfloor n/2 \rfloor}^{\ddagger}(a,1),
\end{equation}
by the construction in figure \ref{figure3}.  Hence
\begin{equation}
\liminf_{n\to\infty} \sfrac{1}{n} \log W_n(a,y) \ge \kappa (a) = \psi(a,1)
\label{eqn:liminfW}
\end{equation}
for all $a > 0$.  Then (\ref{eqn:limsupW}) and (\ref{eqn:liminfW}) complete the proof.
\qed

In particular, when $0 < a \le a_c$ and $y \le 1$ the free energy is equal to $\log \mu_d$.  This
is the \emph{free phase}.

\section{Walks repelled from the surface}
\label{sec:SAWrepel}

The case $0<a \le 1$ and $0<y \le1$ in Section \ref{sec:SAWpushed} is of walks
with midpoint pushed towards the surface.  We now look at the case $0 < a \le 1$ 
and $y \ge 1$ (when walks repelled from the surface are also pulled at their midpoint
from the surface).

The idea in this section is to relate walks with any number of visits to walks
with no visits by translating the walk a unit distance in the $x_d$-direction, and 
adding an edge to reconnect it to the origin.  Since we are
pulling at the mid-point there is a complication in that we want the two subwalks (that meet
at the midpoint) to be of equal length so we have to add an additional edge.  This can be 
conveniently accomplished if we work with unfolded walks.

\begin{figure}[t]
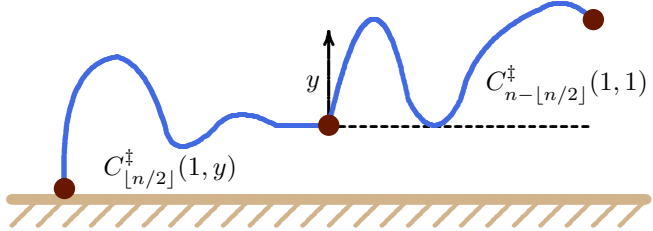

\input figure4.tex
\caption{Concatenating two unfolded walks, the first pulled at its endpoint,
gives a lower bound on $W_n(1,y)$ (that is, the partition function of
a walk pulled in its midpoint).}
\label{figure4}
\end{figure}

We first look at the case $a=1$ where there is no interaction with the surface.
\begin{theo}
When $y \ge 1$ the limit 
$$\lim_{n\to\infty} \sfrac{1}{n} \log W_n(1,y) \equiv \phi(1,y)$$
exists and $\phi(1,y) = \shalf [\lambda(y) + \log \mu_d ].$
\end{theo}
\Pr
We can get an upper bound by regarding the two subwalks that meet at the midpoint as being independent 
and allowing the second sub-walk to penetrate the surface.  This gives
\begin{equation}
\sum_v w_n(v,h) \le \sum_v c_{\lfloor n/2 \rfloor}^+ (v,h) c_{n-\lfloor n/2 \rfloor}
\end{equation}
where $c_m$ is the number of self-avoiding walks with $m$ edges.  Multiplying by $y^h$, summing over $h$,
taking logarithms and dividing by $n$ gives 
\begin{equation}
\limsup_{n\to \infty} \sfrac{1}{n} \log W_n(1,y) \le \shalf[\lambda(y) + \log \mu_d].
\end{equation}
To get a lower bound we work with unfolded walks (see figure \ref{figure4}).  
If we concatenate an unfolded walk pulled 
at its end-point (with $\lfloor \shalf n \rfloor$ edges) and an unfolded positive walk (with 
$n - \lfloor \sfrac{1}{2}n \rfloor$ edges) we have a subset of walks pulled at their mid-point so
\begin{equation}
W_n(1,y) \ge C_{\lfloor n/2 \rfloor}^{\ddagger} (1,y) C_{n-\lfloor n/2 \rfloor}^{\ddagger} (1,1)
\end{equation}
and therefore 
\begin{equation}
\liminf_{n\to \infty} \sfrac{1}{n} \log W_n(1,y) \ge \shalf [\lambda(y) + \log \mu_d]
\end{equation}
which completes the proof.
\qed

\begin{theo}
When $a \le 1$ and $y \ge 1$ the free energy of walks pulled at their mid-point is independent
of $a$.  That is $\lim_{n\to\infty} \sfrac{1}{n} \log W_n(a,y) \equiv \phi(a,y) = \phi(1,y)$ for all $a \le 1$. 
\end{theo}
\Pr
Fix $a \le 1$.  By monotonicity 
\begin{equation}
W_n(0,y) \le W_n(a,y) \le W_n(1,y).
\end{equation}  
Consider walks pulled at their mid-point but unfolded in the $x_1$-direction.
Translate the walk unit distance in the positive $x_d$-direction, add an edge to reconnect to the 
origin and add an edge to the other end of the walk in the positive $x_1$-direction.  The resulting 
walk has no visits and the procedure can be reversed.  In addition the height of the mid-point
changes by 1.  Hence 
\begin{equation}
W_n^{\ddagger}(1,y) = y^{-1}W_{n+2}^{\ddagger}(0,y).
\end{equation}
Then 
\begin{equation}
yW_{n-2}^{\ddagger}(1,y)=W_n^{\ddagger}(0,y) \le W_n(0,y) \le W_n(a,y) \le W_n(1,y)
\end{equation}
and $W_n^{\ddagger}(1,y) \le W_n(1,y) \le W_n^{\ddagger}(1,y)e^{O(\sqrt{n})}$
and the theorem follows.
\qed

\begin{figure}[t]
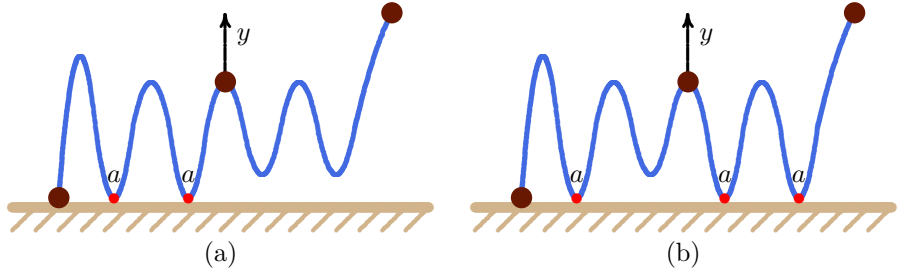

\input figure5.tex
\caption{(a) An adsorbing walk pulled in its mid-point by a vertical force 
with last visit \textit{before} the mid-point.
(b) An adsorbing walk with visits \textit{after} the mid-point (where it is pulled).}
\label{figure5}
\end{figure}

\section{Desorbing a self-avoiding walk by applying a force at the middle vertex}
\label{sec:SAWpull}
In this section we shall be primarily concerned with the case $a \ge a_c$ and $y \ge 1$.
We need a preliminary lemma.
\begin{lemm}
When a loop does not interact with the surface and is pulled in its highest plane
$$\lim_{n\to\infty} \sfrac{1}{n} \log L_n(1,y) = \lambda(\sqrt{y}).$$
\end{lemm}
\Pr
Consider a loop with $n$ edges.  Let $m$ be the last vertex of the loop in its top plane
(\emph{i.e.} with largest $x_d$-coordinate).  Reflect the subwalk from the $m$th to the $n$th vertex in 
this plane to give a positive walk with its last vertex in its top plane.  The height of this subwalk
is twice the height of the original loop.  This gives the inequality $\sum_v l_n(v,h) \le \sum_v c_n^+(v,2h)$ 
and consequently
\begin{equation}
L_n(1,y)  \le  \sum_{v,h} c_n^+(v,2h) y^h 
 \le  \sum_{v,h} c_n^+(v,h) (\sqrt{y})^h = C_n^+(1,\sqrt{y}),
 \label{eqn:loopupperbound}
\end{equation}
and therefore 
\begin{equation}
\limsup_{n\to\infty} \sfrac{1}{n} \log L_n(1,y) \le \lambda(\sqrt y).
\label{eqn:looplimsup}
\end{equation}
To obtain a suitable lower bound we shall construct loops from pairs of unfolded bridges
with the same height (which is also their span in the $x_d$-direction).  With $y$ fixed suppose
that $h^*$ is the value of $h$ such that $b_n^{\ddagger}(h^*)y^{h^*} \ge b_n^{\ddagger}(h)y^h$ for all $h$.  
(Note that $h^*$ depends on both $n$ and $y$.)  Then
\begin{equation}
\frac{B_n^{\ddagger}(y)}{n} \le b_n^{\ddagger}(h^*)y^{h^*} \le B_n^{\ddagger}(y)
\end{equation}
and therefore 
\begin{equation}
\lim_{n\to\infty} \sfrac{1}{n} \log b_n^{\ddagger}(h^*)y^{h^*} = \lambda(y).
\end{equation}
Now concatenate an unfolded bridge with $n$ edges and height $h^*$ with another 
bridge, reflected in $x_1=0$ and translated,
also with $n $ edges and height $h^*$.  The resulting object is a 
loop with $2n $ edges and span $h^*$.  Hence
\begin{equation}
\left(b_{n}^{\ddagger}(h^*)\right)^2 y^{h^*} = \left(b_{n}^{\ddagger} (h^*) \sqrt{y}^{h^*} \right)^2 \le L_{2n}(1,y).
\end{equation}
Taking logarithms, dividing by $2n$ and letting $n \to \infty$ gives
\begin{equation}
\lambda (\sqrt y) \le \liminf_{n\to\infty} \sfrac{1}{n} \log L_n(1,y).
\label{eqn:loopliminf}
\end{equation}
Then (\ref{eqn:looplimsup}) and (\ref{eqn:loopliminf}) complete the proof.
\qed  

This result will be used in the main theorem of this section.

\begin{remark}
Essentially the same proof can be used to show that loops pulled at their mid-point, and 
that loops that have their mid-point
in the top plane and are pulled at this mid-point, also have free energy equal to $\lambda(\sqrt{y})$.
\end{remark}

\begin{theo}
When $a \ge 1$ and $y \ge 1$
$$\phi(a,y)=\lim_{n\to\infty} \sfrac{1}{n} \log W_n(a,y) = \max [\kappa(a), 
\shalf\!\left(\lambda(y) + \log \mu_d\right)].$$
\end{theo}
\Pr
Fix $a \ge 1$ and $y \ge 1$.  By monotonicity $W_n(a,y) \ge \max [ W_n(a,1), W_n(1,y)]$ so
\begin{equation}
\liminf_{n\to\infty} \sfrac{1}{n} \log W_n(a,y) \ge \max [\kappa (a), \shalf(\lambda(y) + \log \mu_d)].
\label{eqn:b0}
\end{equation}
For a walk pulled at its mid-point either the last visit to the surface is before (or at) the mid-point,
or it is after the mid-point (see figure \ref{figure5}).  
If the last visit is before (or at) the mid-point (case (a) in figure \ref{figure5}), an upper bound on 
the partition function of these walks is obtained by cutting the walk in its mid-point
into an adsorbing walk of length $\lfloor \sfrac{1}{2} n \rfloor$ pulled at its endpoint,
and a walk of of length $n - \lfloor \sfrac{1}{2} n \rfloor$.  This gives the upper
bound $C_{\lfloor n/2 \rfloor}^+(a,y) \, c_{n-\lfloor n/2 \rfloor}$ and 
\begin{eqnarray}
\fl \quad \quad
\lim_{n \to \infty} \sfrac{1}{n} \log (C_{\lfloor n/2 \rfloor}^+(a,y)\, c_{n-\lfloor n/2 \rfloor})
 & = &  \shalf\!\left(
\max[\kappa(a), \lambda(y)] +  \log \mu_d \right)  \nonumber \\
& = & \max [ \shalf (\kappa(a) + \log \mu_d), \shalf (\lambda(y) + \log \mu_d) ].
\label{eqn:b1}
\end{eqnarray}
The other case (case (b) in figure \ref{figure5}) 
is where the last visit is after the mid-point of the walk.  The 
middle vertex where the walk is pulled is in a loop that 
only has its first and last vertices in $x_d=0$.  This partitions the walk into 
three subwalks: 
\begin{enumerate}
\item a positive walk interacting with the surface that starts and end in the surface, 
\item a loop with only the first and last vertices in the surface and subject to a force, and 
\item a positive walk interacting with the surface but with no force.  
\end{enumerate}
Note that the loop containing the mid-point has vertical span at least as large as 
the height of the mid-point.  If these three subwalks are treated independently
we have the following upper bound on the partition function:
\begin{eqnarray}
& 
ay \sum_{3 \leq m_2 \le n} \sum_{0 \le m_1 \le m_2-3}
C_{m_1}^+(a,1) \, L_{m_2-m_1-2} (1,y)\, C_{n-m_2}^+(a,1) \nonumber \\
& \le  ay\,n^2 \max_{m_1, m_2}
[e^{\kappa (a) m_1} e^{\lambda (\sqrt {y}) (m_2-m_1-2)} e^{\kappa (a) (n-m_2)}e^{o(n)}]
\nonumber \\
& =  ay \, n^2 \max_m [e^{\kappa(a)(n-m)+\lambda(\sqrt{y})(m-2) + o(n)}].
\end{eqnarray}

\begin{figure}[t]
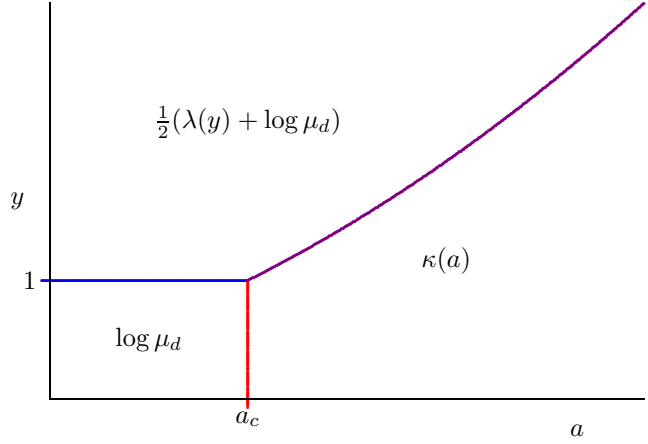

\input figure6.tex
\caption{The phase diagram of adsorbing walks pulled at their midpoint.  There are
three phases:  A free phase when $a<a_c$ and $y < 1$, a ballistic phase when
$\phi(a,y) = \sfrac{1}{2}(\lambda(y) + \log \mu_d)$, and an adsorbed phase
when $\phi(a,y) = \kappa(a)$.  The phase boundary between the adsorbed and 
ballistic phases is given by the solution of 
$\kappa(a) = \sfrac{1}{2}(\lambda(y) + \log\mu_d)$.}
\label{figure6}
\end{figure}
\noindent
Thus, it follows that
\begin{eqnarray}
& \limsup_{n\to\infty} \sfrac{1}{n} \log \left( ay 
\sum_{3 \le m_2 \le n} \sum_{0 \le m_1 \le m_2-3}
C_{m_1}^+(a,1) \, L_{m_2-m_1-2} (1,y)\, C_{n-m_2}^+(a,1) \right) \nonumber \\
& \le \max[\kappa(a),\lambda(\sqrt{y})].
\label{eqn:b2}
\end{eqnarray}
Equations (\ref{eqn:b1}) and (\ref{eqn:b2}) then imply that
\begin{equation}
\limsup_{n\to\infty} \sfrac{1}{n} \log W_n(a,y) \le \max[\kappa(a), \lambda(\sqrt{y}), 
\shalf\!\left(\lambda(y) + \log \mu_d \right)]
\label{eqn:b3}
\end{equation}
since $\kappa(a) \ge \log \mu_d$.  Since $\lambda(y)$ is a convex function of $\log y$ 
\begin{equation}
\lambda(\sqrt{y}) \le \shalf [\lambda(y) + \log \mu_d].
\label{eqn:b4}
\end{equation}
Then equations (\ref{eqn:b0}), (\ref{eqn:b3}) and (\ref{eqn:b4}) complete the proof.
\qed

The phase boundary between the ballistic and adsorbed phase is the locus of the solution of the 
equation $\kappa(a) = \shalf [\lambda(y) + \log \mu_d]$ for $a> a_c$ and $y>1$.  The argument given in 
\cite{Guttmann2014} works \emph{mutatis mutandis} to prove that this phase transition is first order.

These results, taken together, give considerable information about the form of the phase 
diagram in the $(a,y)$-plane and we give a sketch in figure \ref{figure6}.

\section{Low temperature asymptotics}
\label{sec:lowT}
The results of Section \ref{sec:SAWpull} show that the phase boundary between the adsorbed and 
ballistic phases is given by the solution of the equation $\kappa (a) = \shalf [\lambda (y) + \log \mu_d]$.
We can say something useful about the low temperature limit because
we know the behaviour of $\kappa(a)$ and $\lambda(y)$ when $a$ and $y$ are
large \cite{Rensburg2013,RychlewskiJSP}.  In fact $\kappa(a)$ is asymptotic to $\log a + \log \mu_{d-1}$
as $a \to \infty$ \cite{RychlewskiJSP} and $\lambda(y)$ is asymptotic to $\log y$ as $y \to
\infty$ \cite{Rensburg2013}.  Recalling that $a = \exp[-\epsilon/k_BT]$ and $y=\exp[f/k_BT]$ this gives
\begin{equation}
f_c(T) \to  -2 \epsilon + [2 \log \mu_{d-1} - \log \mu_d] k_BT
\end{equation}
as $T \to 0$.  At $T=0$ the required force is twice as large as the force needed when the walk
is pulled at its last vertex \cite{Rensburg2013}.  When $d=3$ $\mu_3$ is about 4.68 and $\mu_2$
is about 2.638 \cite{MadrasSlade} so $\lim_{T\to 0} df_c(T)/dT > 0$ and the force-temperature
curve is re-entrant.  When $d=2$ $\mu_1=1$ so $\lim_{T\to 0} df_c(T)/dT < 0$ because the walk gains 
entropy in the ballistic phase.  Compare this with the case of pulling at the last vertex \cite{Rensburg2013}
where $\lim_{T\to 0} df_c(T)/dT = 0$ when $d=2$.

\section{Pulling at other interior vertices}
\label{sec:SAWinterior}
In this section we consider pulling at an interior vertex other than the middle vertex.

Suppose that we have a positive walk with $n$ edges and we pull at the vertex labelled 
$m=\lfloor \alpha n \rfloor$, $0 < \alpha < 1$.  Let $w_n^{\alpha}(v,h)$ be the number of positive walks
with $n$ edges, with $v+1$ vertices in $x_d=0$ and with $x_d$-coordinate of the $m$th vertex equal to $h$.
Define the partition function $W_n^{\alpha}(a,y) = \sum_{v,h} w_n^{\alpha}(v,h) a^v y^h$.
Clearly $\lim_{n\to\infty} \sfrac{1}{n} \log W_n^{\alpha}(a,1) = \kappa(a)$.  

The arguments developed in Sections \ref{sec:SAWpushed} and \ref{sec:SAWrepel} generalize
easily to the case of a walk pulled at any interior vertex, $0 < \alpha < 1$.  For $a \le a_c$ 
and $y \le 1$ the free energy
is equal to $\log \mu_d$ and the system is in the \emph{free phase}.  When $y \le 1$ the free energy
is $\kappa(a)$, independent of $y$ and when $a \le 1$ and $y \ge 1$ the free energy is 
equal to $\alpha \lambda(y) + (1-\alpha) \log \mu_d$, independent of $a$.

Fix $a \ge 1$ and $y \ge 1$.  By repeating the argument in Section \ref{sec:SAWpull} for
the case $\alpha = \shalf$ it is easy to see that
\begin{equation}
\liminf_{n\to\infty} \sfrac{1}{n} \log W_n^{\alpha} (a,y) \ge \max [\kappa (a), \alpha \lambda(y) + (1-\alpha) \log \mu_d].
\label{eqn:alphalowerbound}
\end{equation}
Similarly we can derive the corresponding upper bound
\begin{equation}
\limsup_{n\to\infty} \sfrac{1}{n} \log W_n^{\alpha}(a,y) \le \max[\kappa (a), \lambda(\sqrt{y}),\alpha \lambda(y) +
(1-\alpha) \log \mu_d].
\label{eqn:alphaupperbound}
\end{equation}
If $\shalf \le \alpha \le 1$ then $\lambda(\sqrt{y}) \le \lambda(y^{\alpha})$ and, since $\lambda(y)$
is a convex function of $\log y$, $\lambda(y^{\alpha}) \le \alpha \lambda(y) + (1-\alpha) \log \mu_d$.
Consequently, using (\ref{eqn:alphalowerbound}) and (\ref{eqn:alphaupperbound}),
\begin{equation}
\lim_{n\to\infty} \sfrac{1}{n} \log W_n^{\alpha}(a,y) = \max[\kappa(a), \alpha \lambda(y) + (1-\alpha) \log \mu_d]
\end{equation}
for all $\alpha \ge \shalf$ for $a \ge 1$ and $y \ge 1$.

This gives a complete description of the phase diagram when $\alpha > \shalf$.  When $\alpha < \shalf$
our results are less complete but there are interesting differences.  The key distinction in this situation is that
a walk that is extended as far as possible by the applied force can still return to the adsorbing
plane.

\begin{figure}[t]
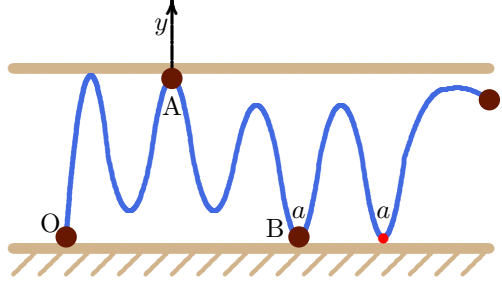

\input figure7.tex
\caption{A schematic diagram of an $LA$-walk.  The pulling force is applied at a vertex marked
A in the top plane of the loop from O to B and is a distance $\lfloor \alpha n \rfloor$ along the walk from O.
The walk returns for the first time to the adsorbing plane at B, and the length of the walk
from O to B is $2\lfloor \alpha n \rfloor$.  The remaining part of the walk, from B to its endpoint,
is an adsorbing walk which is not directly affected by the pulling force at A.  OAB is 
a loop of length $2\lfloor \alpha n \rfloor$ pulled in its midpoint which is also in the top plane of the loop.}
\label{figure7}
\end{figure}

When $\alpha < \shalf$ we shall proceed by constructing a strategy lower bound on the 
partition function.  The idea is to consider the subset of walks where the walk leaves the 
surface at its first step and returns for the first time at vertex $2 \lfloor \alpha n \rfloor$.  Vertex
$\lfloor \alpha n \rfloor$ is in the top plane of the loop from the origin to vertex 
$2 \lfloor \alpha n \rfloor$.  (Note that, by this definition, the vertex at which the force is applied
is in the top plane of the loop.)  We shall call these walks \emph{LA-walks} to recall that the first part 
is a loop pulled at its midpoint and the remainder is a walk that can adsorb with no force.  For a sketch of an
LA-walk see figure \ref{figure7}.
Suppose that the partition function of these walks is ${\cal{L}}_n(a,y,\alpha)$.

\begin{lemm}
The free energy of LA-walks is given by
$$\chi_{LA}(a,y,\alpha)=\lim_{n\to\infty} \sfrac{1}{n} \log {\cal{L}}_n(a,y,\alpha)  =
2 \alpha \lambda(\sqrt{y}) + (1- 2\alpha) \kappa(a).$$
\end{lemm}
\Pr
The partition function of loops with $2 \lfloor \alpha n \rfloor$ edges that have only their first and last vertices
in the adsorbing surface, pulled in their top plane, is $y L_{2 \lfloor \alpha n \rfloor -2} (1,y)$.  Concatenate these with 
positive walks with $n - 2 \lfloor \alpha n \rfloor $ edges giving the upper bound 
\begin{equation}
\limsup_{n\to\infty} \sfrac{1}{n} \log {\cal{L}}_n(a,y,\alpha) \le 2 \alpha \lambda(\sqrt{y}) + (1-2\alpha) \kappa(a).
\end{equation}
We construct a lower bound by concatenating unfolded loops (pulled at their mid-point that is conditioned to be 
in their top plane) with unfolded positive walks, 
with an intermediate edge.  The free energy of these loops is $\lambda(\sqrt{y})$ (see Remark 1).  This gives 
\begin{equation}
\liminf_{n\to\infty} \sfrac{1}{n} \log {\cal{L}}_n(a,y,\alpha) \ge 2 \alpha \lambda(\sqrt{y}) + (1-2\alpha) \kappa(a)
\end{equation}
and these two bounds complete the proof.
\qed

We shall now use this result to show that there are regions of the $(a,y)$-plane where the free
energy is greater than $\max [\kappa(a), \alpha \lambda(y) + (1-\alpha) \log \mu_d]$.  We first 
observe that $\chi_{LA} > \kappa(a)$ if and only if $\lambda(\sqrt{y}) > \kappa(a)$.  Now 
$\lambda(\sqrt{y}) \ge \shalf \log y $ and $\kappa(a) \le \log a + \log \mu_d$ so if 
\begin{equation}
\log y > 2\log a + 2\log \mu_d
\label{eqn:condition1}
\end{equation}
then $\lambda(\sqrt{y}) > \kappa(a)$.

Since $\lambda(y) \le \log y + \log \mu_d$ and $\kappa(a) \ge \log a + \log \mu_{d-1}$ 
(see for instance \cite{HTW}) we
observe that the condition
\begin{equation}
\log a > \frac{\log \mu_d}{1-2\alpha} - \log \mu_{d-1}
\label{eqn:condition2}
\end{equation}
implies that 
\begin{equation}
\chi_{LA} > \alpha \lambda(y) + (1-\alpha) \log \mu_d.
\end{equation}
Hence if conditions (\ref{eqn:condition1}) and (\ref{eqn:condition2}) are both satisfied then we are assured
that the free energy is larger than $\max [\kappa(a), \alpha \lambda(y) + (1-\alpha) \log \mu_d]$
and there is an additional phase in the phase diagram.  For any $0 < \alpha < \shalf$ both 
conditions can always be satisfied by making $a$ and $y$ sufficiently large.  For instance, 
if $\alpha = \squarter$ then sufficient conditions are $\log a > 2 \log \mu_d - \log \mu_{d-1}$ and
$\log y > 2 \log a + 2 \log \mu_d$.

\section{Discussion}
\label{sec:discussion}
Earlier work has focused on pulling a terminally attached  self-avoiding walk from a surface
at which it is adsorbed 
by applying a force at the last vertex of the 
walk \cite{Guttmann2014,Rensburg2013,Krawczyk2005,Krawczyk2004,Mishra2005}, 
or in the plane containing the vertices furthest
from the surface \cite{Rensburg2016b}.  From the experimental point 
of view there are interesting questions about how the behaviour depends on where the force is 
applied and, in this paper, we consider the case where the force is applied (normal to the surface)
at the mid-point of the walk.  We show that the phase diagram in the $(a,y)$-plane
is qualitatively similar to that for the case where the force is applied at the last vertex but 
the phase boundary between the adsorbed and ballistic phases is shifted.  That is, the critical force 
required for desorption depends on where the force is applied.  When we switch to the force-temperature
plane there are distinct differences in the low temperature behaviour depending on where the force is 
applied.

We have also considered the case where the force is applied at an interior vertex other than the 
middle vertex.  Our results are less complete but we have shown that, in some circumstances, 
the critical force for desorption changes when we change the vertex at which the force is applied.
When the force is applied between the middle vertex and the free vertex of degree 1 (not attached to the surface)
the results depend on the particular vertex at which the force is applied, but the transition is from an 
adsorbed to a ballistic phase, as in the case when the force is applied at the middle vertex.
When the force is applied between the middle vertex and the point of attachment we have shown
that there is an intermediate phase for some values of $a$ and $y$ and we 
have bounds on these values.  In this phase 
we have a lower bound on the free energy that should be especially effective at large 
$a$ and $y$ but it is unlikely that this bound will be strict 
throughout this phase.  The walks in this phase are expected to consist of a loop that is 
extended by the force but the walk then returns to the surface and the remainder of the 
walk has a positive density of visits.  LA-walks are a subset of these walks.

\section*{Acknowledgement}
This work began as a result of a question asked by Professor T. Cosgrove and we thank him
for his input.  This research was partially supported by NSERC of Canada.

\end{document}